\documentclass[conference,a4paper]{IEEEtran}

\DeclareRobustCommand*{\IEEEauthorrefmark}[1]{\raisebox{0pt}[0pt][0pt]{\textsuperscript{\footnotesize #1}}}

\usepackage{cite}
\usepackage{amsmath,amssymb,amsfonts}
\usepackage{algorithmic}
\usepackage{graphicx}
\usepackage{textcomp}
\usepackage{xcolor}
\usepackage{hyperref}

\begin{document}
%

\title{Modeling flux tunability in Josephson Traveling Wave Parametric Amplifiers with an open-source frequency-domain simulator}

\author{\IEEEauthorblockN{
A. Levochkina*\IEEEauthorrefmark{1,2},   
I. Chatterjee\IEEEauthorrefmark{1,2},    
P. Darvehi\IEEEauthorrefmark{2},     
H. G. Ahmad\IEEEauthorrefmark{1,2}, 
P. Mastrovito\IEEEauthorrefmark{1,2},
D. Massarotti\IEEEauthorrefmark{3,2},\\
D. Montemurro\IEEEauthorrefmark{1,2},
F. Tafuri\IEEEauthorrefmark{1,4},
G.P. Pepe\IEEEauthorrefmark{1,2},
Kevin P. O’Brien\IEEEauthorrefmark{5} and
M. Esposito\IEEEauthorrefmark{2},
}                                     
\IEEEauthorblockA{\IEEEauthorrefmark{1}
Department of Physics, University Federico II, Naples, 80126, Italy}
\IEEEauthorblockA{\IEEEauthorrefmark{2}
CNR-SPIN, Complesso di Monte S. Angelo, via Cintia, Napoli, 80126, Italy}
\IEEEauthorblockA{\IEEEauthorrefmark{3}
Department of Electrical Engineering and Information Technology, University
Federico II, Naples, 80125, Italy}
\IEEEauthorblockA{\IEEEauthorrefmark{4}
CNR INO, Largo Enrico Fermi 6, Florence, 50125, Italy,} 
\IEEEauthorblockA{\IEEEauthorrefmark{5}
Department of Electrical Engineering and Computer Science,
 Massachusetts Institute of Technology,\\
 Cambridge, MA, United States
}
* anna.levochkina@unina.it}

\maketitle

\begin{abstract}
Josephson Traveling Wave Parametric Amplifiers (JTWPAs) are integral parts of many experiments carried out in quantum technologies. Being composed of hundreds of Josephson junction-based unit cells, such devices exhibit complex nonlinear behavior that typically cannot be fully explained with simple analytical models, thus necessitating the use of numerical simulators. A very useful characteristic of JTWPAs is the possibility of being biased by an external magnetic flux, allowing in-situ control of the nonlinearity. It is therefore very desirable for numerical simulators to support this feature. Open-source numerical tools that allow to model JTWPA flux biasing, such as \textit{WRSPICE} or \textit{PSCAN2}, are based on time-domain approaches, which typically require long simulation times to get accurate results. In this work, we model the gain performance in a prototypical flux-tunable JTWPA by using \textit{JosephsonCircuits.jl}, a recently developed frequency-domain open-source numerical simulator, which has the benefit of simulation times about 10,000 faster than time-domain methods. By comparing the numerical and experimental results, we validate this approach for modeling the flux dependent behavior of JTWPAs.
\end{abstract}

{\smallskip \keywords Josephson Traveling Wave Parametric Amplifier, JTWPA, SNAIL TWPA,  Julia, JosephsonCircuits.jl }

\IEEEpeerreviewmaketitle

\section{Introduction}

After the first experimental demonstration of a near quantum limited Josephson Traveling Wave Parametric Amplifier (JTWPA) in 2015 \cite{Macklin15}, it soon became clear that exploiting the nonlinearity of long chains of Josephson junctions is a powerful approach to obtain low noise broadband amplification in the microwave regime. Since then, this research field has grown very quickly, and impressive progress has been made \cite{Aumentado20, Esposito21, malnou24}. Still, many challenges remain open in order to improve JTWPAs performance and tailor these devices for diverse quantum technology applications, for example, dark matter search \cite{Braggio22, bartram23, Di_vora23} or squeezing generation \cite{Esposito22, Perelshtein22, Qiu23, Casariego_2023}.

The simplest possible JTWPA unit cell contains a single Josephson junction, without flux tunability, as for the device in \cite{Macklin15}. As research progressed,  more complex unit cells were introduced by embedding one or more Josephson junctions in a superconducting loop and allowing flux tunability. Examples are JTWPAs based on SQUIDs (Superconducting Quantum Interference Devices \cite{barone82}) \cite{Zorin16, Bell15, Zorin19, Fasolo24} as well as SNAILs (Superconducting Nonlinear Asymmetric Inductive eLements \cite{Frattini18}) \cite{Ranadive22, Roudsari23}. Flux-biasing in JTWPAs has the advantage of providing in-situ control on the nonlinearity. This allows to investigate different amplification mechanisms, such as \textit{three wave mixing} \cite{Zorin16}, \textit{reversed Kerr} \cite{Ranadive22} and \textit{dynamical phase matching} \cite{Ranadive24} amplification.



Realistically modeling the gain performance of JTWPAs often requires the use of numerical circuit simulations, which improve the prediction capabilities in comparison to approximated analytical approaches \cite{Dixon20, Peng22, gaydamachenko_numerical_2022, Kissling23, Peatain23, levochkina_numerical_2024, Levochkina24}. 
Software like JSIM \cite{JSIM}, WRspice \cite{WRSPICE}, PSCAN \cite{PSCAN}, and PSCAN2 \cite{PSCAN2} enable the time-domain simulation of superconducting circuits with Josephson junctions and are often used for modeling JTWPAs via the Fourier analysis of the time-domain results. 
They offer various methods for the simulation of flux-biased devices, but they are limited by the long simulation time required to get accurate results for large circuits like JTWPAs 
\cite{levochkina_numerical_2024}. 


Frequency-domain approaches such as the harmonic balance method are a more computationally efficient alternative \cite{Peng22, levochkina_numerical_2024,model_flux_24}.The recently released open source JosephsonCircuits.jl package \cite{JosephsonCiruits.jl} enables quick frequency-domain based simulations of JTWPAs and has been initially validated for JTWPA designs without flux tunability \cite{Peng22}. 
A technique for modeling JTWPA flux biasing has been recently added to the JosephsonCircuits.jl documentation \cite{JosephsonCiruits.jl}, considering a specific example of a three wave mixing (3WM) flux-pumped device with SQUID-based unit cells \cite{Zorin19}. 
%

By appropriately adapting the example provided in the software documentation, we tested the JosephsonCircuits.jl simulator for predicting flux-tunable amplification in a current-pumped JTWPA. We validated the results by comparing the simulated gain with experimental data for a prototypical flux-tunable JTWPA with SNAIL-based unit cells. 


\section{Prototypical Flux-tunable JTWPA}
We perform experiments with a prototypical JTWPA device consisting of 700 unit cells, each containing a SNAIL (Superconducting Nonlinear Asymmetric Inductive eLement) \cite{Frattini18} and a capacitance to ground. The adopted device is nominally identical to the one in \cite{Ranadive22}. The experimental setup is the same as described in \cite{Levochkina24} for TWPA gain measurements. 

\begin{figure}[!htbp]
\centering
\includegraphics[width=0.35\textwidth]{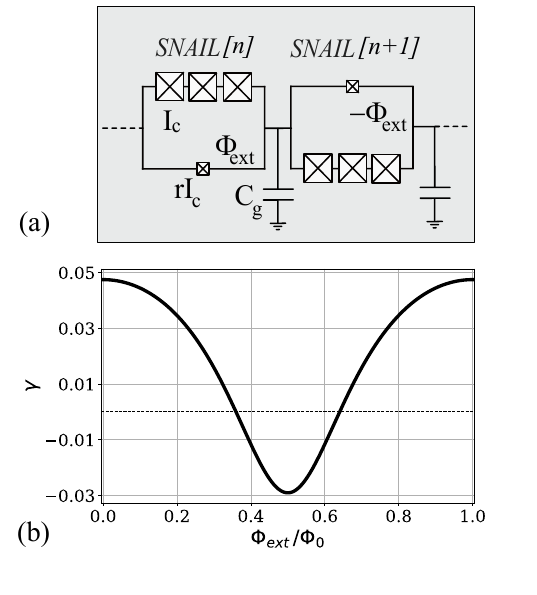}
\caption{(a) Sketch of two adjacent unit cells in a SNAIL-based JTWPA device with alternating flux polarity \cite{Ranadive22}. (b) Four wave mixing coefficient $\gamma$ as a function of normalized external magnetic flux.}
\label{fig:figure1}
\end{figure}

In Fig. \ref{fig:figure1} (a), we show a sketch of two adjacent unit cells in the measured device.
\\

The current-phase relation for each SNAIL can be expressed with the following Taylor expansion:
\begin{equation} \label{eq:1}
    \frac{I_L(\phi^* + \phi)}{\Tilde{\alpha}I_C} \approx 
     \phi - {\beta}(\phi)^2 - {\gamma}(\phi)^3,
\end{equation}
where $I_c$ is the critical current of the big Josephson junctions, and the expansion is about a flux $\phi^*$ such that $I_L(\phi ^*)=0$. The coefficients in Eq. \ref{eq:1} are defined as follows:
\begin{equation} \label{eq:2}
    \Tilde{\alpha} = r \cos \phi^* + \frac{1}{3}\cos\left(\frac{\phi^*-\phi_{\text{ext}}}{3}\right),
\end{equation}
\begin{equation} \label{eq:3}
    {\beta} = \frac{1}{2}\left[r \sin \phi^* + \frac{1}{9} \sin\left(\frac{\phi^*-\phi_{\text{ext}}}{3}\right)\right]/\Tilde{\alpha},
\end{equation}
\begin{equation} \label{eq:4}
    {\gamma} = \frac{1}{6}\left[ r \cos\phi^* + \frac{1}{27} \cos\left(\frac{\phi^*-\phi_{\text{ext}}}{3}\right)\right]/\Tilde{\alpha} \, ,
\end{equation}
where $r$ is the critical current ratio between small and big Josephson junctions, and $\beta$ and $\gamma$ are the nonlinear coefficients associated with three wave mixing (3WM) and four wave mixing (4WM) nonlinear processes, respectively. The flux tunability in this JTWPA comes from the dependence on the external magnetic flux $\Phi_{\text{ext}}$ of the nonlinear coefficients $\beta$ and $\gamma$. 
In experiments, $\Phi_{\text{ext}}$ is easily controlled by sending DC current in a superconducting coil underneath the device. \\
As shown in Fig. \ref{fig:figure1} (a), adjacent SNAILs have opposite orientation, generating magnetic fluxes with opposite signs. Such an alternating flux polarity design \cite{Ranadive22} entails an overall suppression of the $\beta$ coefficient, as the latter is an odd function of the flux. 
%
The flux dependence of the $\gamma$ coefficient for the device under test is reported in Fig.\ref{fig:figure1} (b). 
In the following, we focus on 4WM gain for different values of the external magnetic flux, demonstrating that the adopted open-source frequency-dependent simulator can accurately reproduce the experimental results.

\section{Numerical Simulations with JosephsonCircuits.jl}
\begin{figure}[h]
\includegraphics[width=0.48\textwidth]{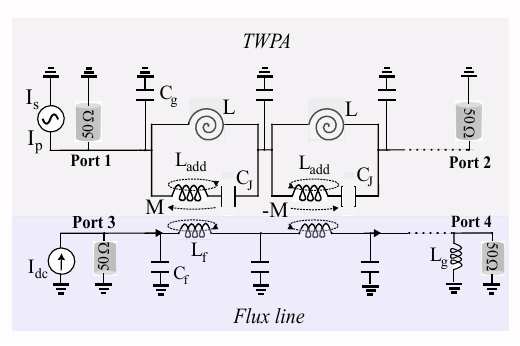}
\caption{Equivalent ciruit of a SNAIL TWPA with 700 SNAILs for JosephsonCircuits.jl simulations. Here we highlight the scheme adopted for simulating the flux-biasing of two adjacent unit cells. We set opposite signs for mutual inductances of adjacent cells to model the alternating flux polarity.}
\label{fig:figure2}
\end{figure}

Using the circuit depicted in Fig. \ref{fig:figure2}, we numerically simulate the JTWPA described in the previous section.
We employ an extra flux line for implementing flux biasing, following the approach recently reported in the open-source simulator documentation \cite{JosephsonCiruits.jl}. 
For each unit cell, the inductance of the additional flux line is mutually coupled with a small linear inductance specifically introduced in the SNAIL circuit for this purpose.
We set mutual inductance with the opposite sign for adjacent SNAILs to simulate the inverted flux polarity design of the considered JTWPA.

Table \ref{Tab1} displays the circuit parameters that are used in the numerical modeling of the device.
\begin{table}[h!]
    \caption{Device Parameters for JosephsonCircuits.jl}   
    \label{Tab1}
    \centering 
     \begin{tabular}{ |p{3cm}||p{3.2cm}| }
     \hline 
     number of cells     & 700  \\
     \hline 
     $I_c$      & 2.19 $\mu$A  \\
     \hline 
     $r$  & 0.07  \\
     \hline 
     $C_J$ & 50 fF\\
     \hline 
     $C_g$   & 250 fF \\
     \hline 
     $L_{add}$  & 70 fH\\
     \hline 
     $L_f$ & 190 pH  \\
     \hline 
     $C_f$ & 0.076 pF  \\
     \hline 
     $L_g$ & 20.0 nH\\     
     \hline
    \end{tabular}
\end{table}
\begin{figure}[ht]
\includegraphics[width=0.48\textwidth]{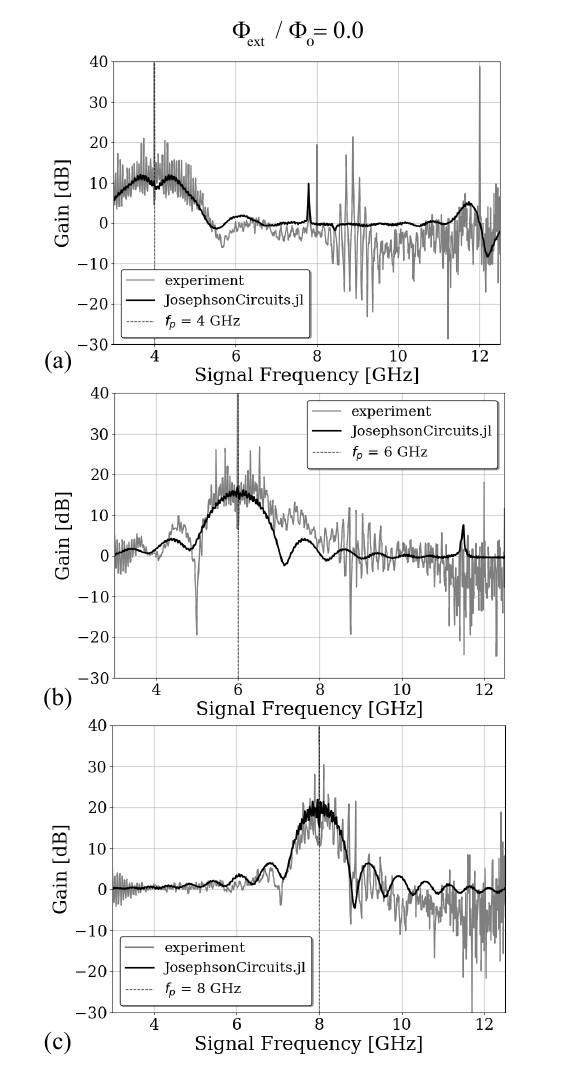}
\caption{Gain comparison for $\Phi_{ext}/\Phi_0$ = 0.0. Grey line refers to the experimentally obtained data,  black line is JosephsonCircuits.jl simulation. (a) pump frequency $f_p$ = 4 GHz, experimental pump power is -72 dBm, simulation pump power is 1.02 $\mu A$ (b) pump frequency $f_p$ = 6 GHz, experimental pump power is -73 dBm, simulation pump power is 0.852 $\mu A$  (c) pump frequency $f_p$ = 8 GHz, experimental pump power is -72 dBm, simulation pump power is 1.02 $\mu A$ }
\label{fig:figure3}
\end{figure}
\begin{figure}[ht]
\includegraphics[width=0.478\textwidth]{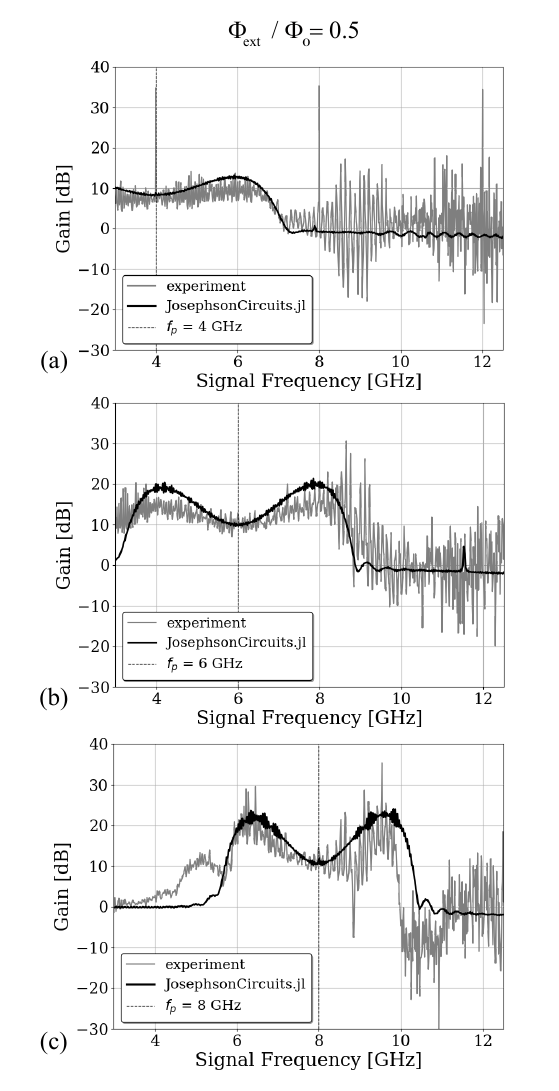}
\caption{Gain comparison for $\Phi_{ext}/\Phi_0$ = 0.5. Grey line refers to the experimentally obtained data,  black line is JosephsonCircuits.jl simulation. (a) pump frequency $f_p$ = 4 GHz, experimental pump power is -72 dBm, simulation pump power is 1.05 $\mu A$ (b) pump frequency $f_p$ = 6 GHz, experimental pump power is -73.5 dBm, simulation pump power is 0.992 $\mu A$  (c) pump frequency $f_p$ = 8 GHz, experimental pump power is -72.5 dBm, simulation pump power is 1.0 $\mu A$ }
\label{fig:figure4}
\end{figure}
Losses in the considered device are primarily arising from the dielectric losses coming from the large capacitance to ground per unit cell, which is required for impedance matching. We consider a loss tangent $\tan(\delta_0)$ = 2.1x$10^{-3}$, estimated from experimental characterizations \cite{Ranadive22}, and include such value in the simulation according to the method described in \cite{JosephsonCiruits.jl}. 

We conduct a comparative study between JTWPA's gain measured experimentally and gain computed with JosephsonCircuits.jl tool for two significant flux points ($\Phi_{ext}/\Phi_0$ = 0.0 and 0.5). Figures \ref{fig:figure3} and \ref{fig:figure4} illustrate the obtained results. 
For both flux points, we consistently find good qualitative agreement between numerical simulations and experiment for three distinct pump frequencies.

For the gain simulation, 8 pump harmonics and 4 harmonic modulation were set. Furthermore, we chose the pump power values in JosephsonCircuits.jl to match the values used in the experiments. External flux $\Phi_{\text{ext}}/\Phi_0$ = 0.5 in the experiment corresponds to $I_{dc}$ = 0.285 $mA$ through the auxiliary flux line in the simulated circuit (Fig. \ref{fig:figure2}). 

A full frequency domain simulation for a total of 523 frequency points takes about 40 seconds to complete on a standard laptop pc.
%
\begin{figure}[t!]
\centering
\includegraphics[width=0.5\textwidth]{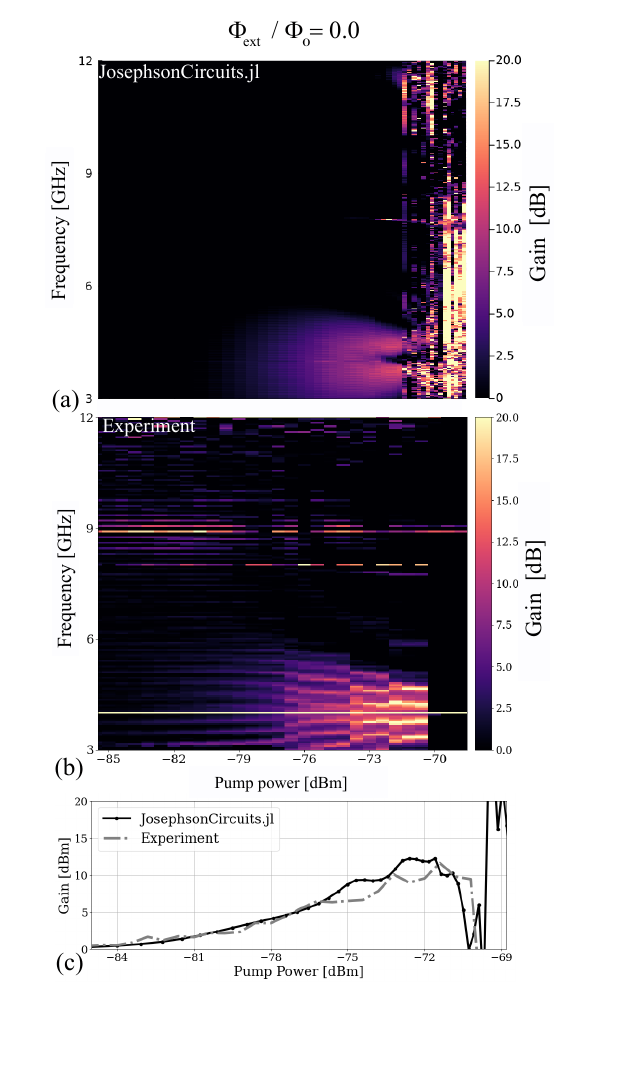}
\caption{Gain as a function of signal frequency and pump power for $\Phi_{ext}/\Phi_0$ = 0.0. Pump frequency $f_p$ = 4 GHz (a) JosephsonCircuits.jl, (b) experimental results. (c) Cut of colormesh plots at signal frequency $f_s$ = 4.4 GHz.}
\label{fig:figure5}
\end{figure}
The addition of the auxiliary flux line, necessary to model the flux tunable gain, increase the simulation time by about 10 seconds compared to case without it \cite{levochkina_numerical_2024}. We stress that such running time is significantly shorter compared to time-domain based software, such as WRspice, which would require more than four days of simulation time to get the same accuracy and frequency resolution \cite{Peng22}.

\begin{figure}[t!]
\centering
\includegraphics[width=0.504\textwidth]{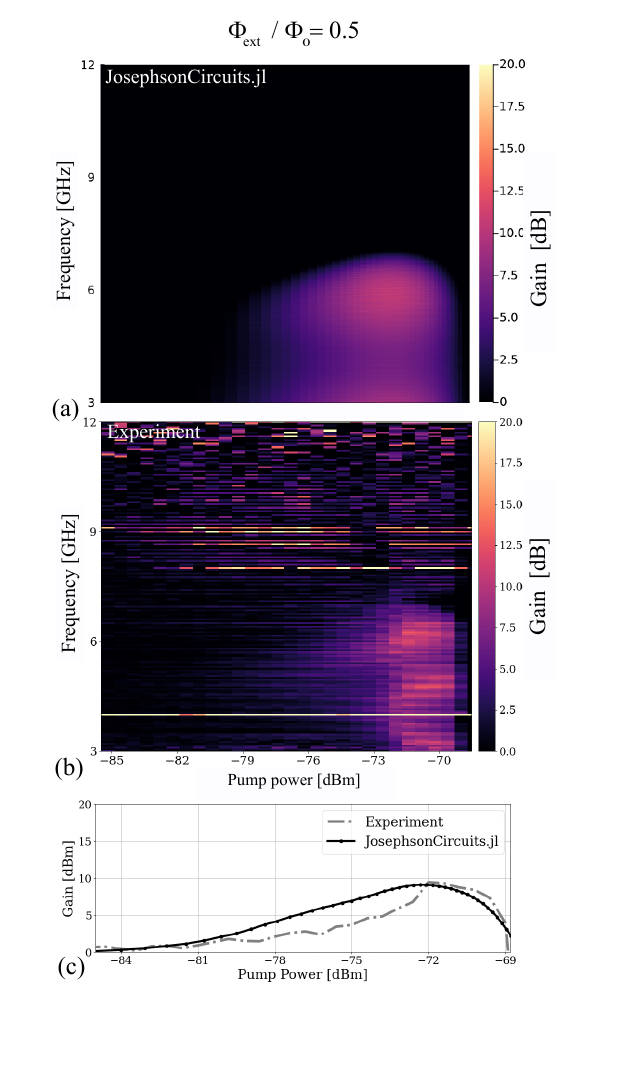}
\caption{Gain as a function of signal frequency and pump power for $\Phi_{\text{ext}}/\Phi_0$ = 0.5. Pump frequency $f_p$ = 4 GHz (a) JosephsonCircuits.jl, (b) experimental results. (c) Cut of colormesh plots at signal frequency $f_s$ = 4.4 GHz.}
\label{fig:figure6}
\end{figure}

Finally, we focus on the gain behavior as a function of the pump power. Fig. \ref{fig:figure5}  and \ref{fig:figure6} report the experimental and simulated gain as a function of the pump power across the entire experimentally available frequency range for the applied flux $\Phi_{ext}/\Phi_0$ = 0.0 and $\Phi_{ext}/\Phi_0$ = 0.5 respectively.
For a direct comparison between models and experiments, horizontal cuts at signal frequency $f_s$ = 4.4 GHz are reported in Figs. \ref{fig:figure5} (c) and \ref{fig:figure6} (c). Also in this case a good qualitative agreement is observed, suggesting the validity of the tested approach.

\section{Conclusion}
In conclusion, we test the accuracy of the recently released open source frequency-domain simulator JosephsonCircuits.jl \cite{JosephsonCiruits.jl} for predicting flux-tunable gain behavior in JTWPAs.
We benchmark this numerical tool in the context of 4WM current-pumped amplification by simulating a real JTWPA device based on SNAILs unit-cells, which has been experimentally characterized in our Lab.

We applied a simulation strategy recently reported in \cite{JosephsonCiruits.jl} for dc-SQUID 3WM flux-pumped JTWPAs and appropriately modified it for the case of a current-pumped 4WM JTWPA with flux tunable non-linearity.  While in the experiment the flux tunability is implemented with an external coil, in the circuit simulation we adopted an auxiliary flux line coupled to the JTWPA circuit.
The device losses were also included into the model.

The match between experiments and simulations demonstrates the accuracy of JosephsonCiruits.jl package for predicting flux-tunable amplification in JTWPAs with the advantage of substantially reduced simulation time with respect to time-domain based simulation approaches. Our results encourage the use of such simulation methods for the design and investigation of novel JTWPA devices.

\section*{Acknowledgments}
This work has been supported by European Union through Horizon Europe 2021-2027 Framework Programme under Grant 101080152 and through Horizon 2020 Research and Innovation Programme under Grant 731473 and Grant 101017733, by PNRR MUR Project PE0000023-NQSTI, and by the MUR Project PRIN 2022 CUP E53D23001910006 (SuperNISQ). We would like to acknowledge A. Ranadive, G. Cappelli, G. Le Gal, N. Roch and L. Planat for helpful discussions. H.G.A. acknowledges support by PNRR MUR Project CN00000013-ICSC. H.G.A, D.M. and F.T. thank the SUPERQUMAP project (COST Action CA21144).

\bibliographystyle{unsrturl}
\bibliography{sample}

\end{document}